# MEGAHIT: An ultra-fast single-node solution for large and complex metagenomics assembly via succinct *de Bruijn* graph

Dinghua Li[1,†], Chi-Man Liu[2,†], Ruibang Luo[2,†], Kunihiko Sadakane[3] and Tak-Wah Lam[1,2,*]

[1] HKU-BGI Bioinformatics Algorithms Research Laboratory & Department of Computer Science, University of Hong Kong, Hong Kong  [2] L3 Bioinformatics Limited, Hong Kong  [3] National Institute of Informatics, Chiyoda-ku, Tokyo, Japan

**ABSTRACT**

**Summary:** MEGAHIT is a NGS *de novo* assembler for assembling large and complex metagenomics data in a time- and cost-efficient manner. It finished assembling a soil metagenomics dataset with 252Gbps in 44.1 hours and 99.6 hours on a single computing node with and without a GPU, respectively. MEGAHIT assembles the data as a whole, i.e., no pre-processing like partitioning and normalization was needed. When compared with previous methods (Chikhi and Rizk, 2012; Howe, et al., 2014) on assembling the soil data, MEGAHIT generated a 3-time larger assembly, with longer contig N50 and average contig length; furthermore, 55.8% of the reads were aligned to the assembly, giving a 4-fold improvement.

**Availability:** The source code of MEGAHIT is freely available at https://github.com/voutcn/megahit under GPLv3 license.

**Contact:** rb@l3-bioinfo.com, twlam@cs.hku.hk

## 1 INTRODUCTION

Next generation sequencing technologies have offered new opportunities to study metagenomics and understand various microbial communities such as human guts, rumen and soil. Due to the lack of reference genomes, *de novo* assembly of metagenomics data (short reads) is a beneficial and almost inevitable step for metagenomics analysis (Qin, et al., 2010). This step is, however, constrained by the heavy requirement of computational resources, especially for large and complex datasets encountered in environmental metagenomics (Howe, et al., 2014). The soil metagenomics dataset recently published by Howe et al. comprises 252G basepairs even after trimming low quality bases. The dataset was successfully assembled with pre-processing steps including partitioning and digital normalization. At present no *de novo* assembler can assemble the data as a whole using a feasible amount of computer memory. Estimated memory requirement for SOAPdenovo2 (Luo, et al., 2012) and IDBA-UD (Peng, et al., 2012) to assemble the soil data is at least 4TB. As the volume of metagenomics data keeps growing, we are motivated to develop MEGAHIT, an assembler that can assemble large and complex metagenomics data in a time- and cost-efficient manner, especially on a single-node server (current maximum memory capacity 768 GB for a 2-socket server).

## 2 METHODS

MEGAHIT makes use of succinct *de Bruijn* graphs (Bowe, et al., 2012), which are compressed representation of *de Bruijn* graphs. A succinct *de Bruijn* graph (SdBG) encodes a graph with $m$ edges in $O(m)$ bits, and supports $O(1)$ time traversal from a vertex to its neighbors. Our implementation has added a bit-vector of length $m$ to mark the validity of each edge (so as to support dynamic removal of edges efficiently), and an auxiliary vector of $2kt$ bits (where $k$ is the k-mer size and $t$ is the number of zero-indegree vertices) to store the sequence of zero-indegree vertices to ensure the graph being lossless.

Despite its advantages, constructing a SdBG efficiently is non-trivial. MEGAHIT is rooted in a fast parallel algorithm for SdBG construction; the bottleneck is sorting a set of $(k+1)$-mers that are the edges of an SdBG in reverse lexicographical order of their length-$k$ prefixes ($k$-mers). MEGAHIT exploits the parallelism of a Graphics Processing Unit (GPU, CUDA-enabled) by adapting the recent BWT-construction algorithm CX1 (Liu, et al., 2014), which takes advantage of a GPU to sort the suffices of a set of reads very efficiently. Limited by the relatively small size of GPU's on-board memory, we adopt a block-wise strategy that partitions the $k$-mers according to their length-$l$ prefix ($l$=8 in our implementation). The $k$-mers in consecutive partitions that fit within the GPU memory are sorted together. Leveraging the parallelism of GPU, MEGAHIT speeds up the construction by 3-5 times over its CPU-only counterpart.

Notably, sequencing error is problematic, because a single base of sequencing error leads to $k$ erroneous $k$-mer singletons, which increases the memory consumption of MEGAHIT significantly. To cope with the problem, before graph construction, all $(k+1)$-mers from the input reads are sorted and counted, and only $(k+1)$-mers that appear at least $d$ (2 by default) times are kept as *solid-kmer*. This method removes many spurious edges, but may be risky for metagenomics assembly since many low-abundance species may have been sequenced at very low depth. Thus we introduce a *mercy-kmer* strategy to recover these low-depth edges. Given two *solid* $(k+1)$-mers $x$ and $y$ from the same read, where $x$ has no outdegree and $y$ has no indegree. If all $(k+1)$-mers between $x$ and $y$ in that read are not *solid*, they will be added to the *de Bruijn* graph as *mercy-kmers*. *Mercy-kmers* strengthen the contiguity of low-depth regions. Without this approach, many authentic low-depth edges would be incorrectly identified as tips and removed.

Based on SdBG, we implemented a multiple $k$-mer size strategy in MEGAHIT (Peng et al., 2012). The method iteratively builds multiple SdBGs from a small $k$ to a large $k$. While a small $k$-mer size is favorable for filtering erroneous edges and filling gaps in low-coverage regions, a large $k$-mer size is useful for resolving repeats. In each iteration, MEGAHIT cleans potentially erroneous edges by removing tips, merging bubbles and removing low local coverage edges. The last approach is especially useful for metagenomics, which suffers from non-uniform sequencing depths. The overall workflow of MEGAHIT is shown in Fig 1.

---

[*]To whom correspondence should be addressed.  [†]Equal first-author.

**Fig. 1.** The workflow of MEGAHIT

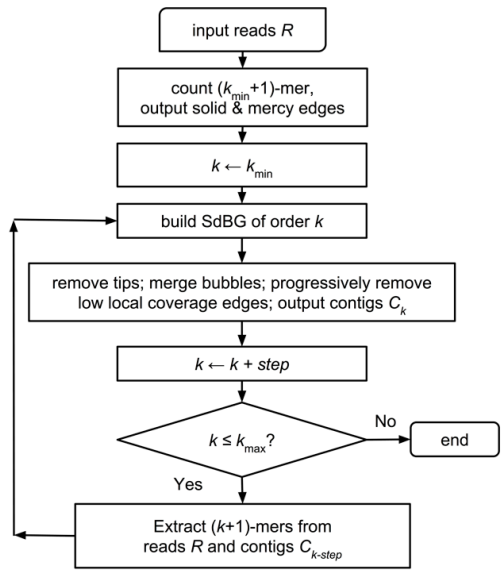

## 3   RESULTS

Table 1 compares the performance of MEGAHIT with SPAdes (Bankevich, et al., 2012) on three subsets (100-fold, 20-fold and 10-fold) of an *E. coli* MG1655 dataset. QUAST (Gurevich, et al., 2013) was used to evaluate the assembled contigs (Table 1). MEGAHIT (CPU version) is 6 times faster than SPAdes, and performs well even on the low-coverage subset.

To evaluate the performance on large scale metagenomics data, we assembled an Iowa prairie soil metagenomics dataset that comprises 3.3 billion reads totaling 252 billion base-pairs (Howe, et al., 2014) using MEGAHIT and Minia, another memory-efficient assembler (Chikhi and Rizk, 2012). The assembly conducted by Howe et al. was included for comparison (Table 2). On a server with 384 GB memory, MEGAHIT took 44.1 hours, ~7 times faster than Minia. It reached peak memory consumption at 345 GB during $k$-mer counting and SdBG construction; this matches the expectation since MEGAHIT's sorting module automatically adjusts to fully utilize all available memory in a server. Notably, MEGAHIT can assemble this dataset with as little as 260GB memory, using 55.3 hours (Supp. Section 4).

To be consistent with Howe's analysis, we only considered contigs ≥300 bp for further analysis. The contigs produced by MEGAHIT had a total size at least 3 times larger than by other methods, and achieved better statistics on N50, average length, and the number of long contigs (length ≥1000bp). Thus MEGAHIT gives better assembly contiguity. Raw reads were aligned back to the assembled contigs using Bowtie2 (Langmead and Salzberg, 2012). As shown in Table 3, MEGAHIT gets >4 times more reads mapped and 5-6 times more read pairs properly aligned. 37% of distinct 17-mers appeared ≥2 in the assembly, which might imply that MEGAHIT did a better job in recovering low-abundance subspecies in ultra-diversified metagenomics (Supp. Fig. S3).

## 4   CONCLUSIONS

MEGAHIT enables an efficient assembly of large and complex metagenomics data on a single server, while giving better completeness and contiguity. MEGAHIT is available in both CPU-only and GPU-accelerated versions. With GPU, the assembly time of the soil dataset is shortened from 4 days to less than 2 days.

**Table 1**. Performance of MEGAHIT and SPAdes on the *E.coli* dataset

|  | MEGAHIT 100x | MEGAHIT 20 x | MEGAHIT 10 x | SPAdes 10 x |
|---|---|---|---|---|
| N50 (bp) | 73,736 | 52,352 | 9,067 | 18,264 |
| Largest Alignment (bp) | 221k | 178k | 31k | 62k |
| bp in contigs >= 1kbp | 4.55M | 4.55M | 4.52M | 4.55M |
| Genome Fraction | 98.0% | 98.1% | 97.4% | 97.9% |
| Misassemblies (bp) | 2k | 41k | 81k | 64k |
| Wall Time (sec.) | 185 | 82 | 47 | 318 |

MEGAHIT: CPU version, options "--k-min 21 --k-max 81 –m 1000000000"; SPAdes and QUAST was run with default parameters.

**Table 2.** Summary statistics for MEGAHIT, Howe et al. and Minia

|  | MEGAHIT | Howe et al. | Minia |
|---|---|---|---|
| Wall Time (hr) | 44.1 | >488 | 331.4 |
| Peak Memory (GB) | 345 | 287 | 29 |
| Total Size (Mbp) | 4,902 | 1,503 | 1,490 |
| Average Length (bp) | 633 | 485 | 505 |
| N50 (bp) | 657 | 471 | 488 |
| Longest (bp) | 184,210 | 9,397 | 32,679 |
| # of Contigs | 7,749,211 | 3,096,464 | 2,951,575 |
| # of Contigs ≥ 1kbp | 841,257 | 129,513 | 158,402 |

MEGAHIT utilizes all 24 CPU threads with options "--k-min 27 --k-max 87 --k-step 10 -m 370000000000". The wall time for CPU version of MEGAHIT is 99.4 hour. Minia does not support multi-threads; it was run with $k$=31 and min_abundance=2. The time and memory of Howe et al. were excerpted from the paper; the time accounts for digital normalization and partitioning only.

**Table 3.** Alignment statistics of MEGAHIT, Howe et al. and Minia

|  | MEGAHIT | Howe et al. | Minia |
|---|---|---|---|
| Total # of reads |  | 3,252,369,195 |  |
| Reads overall aligned (%) | 55.81 | 10.72 | 13.03 |
| Total # of SE reads |  | 356,742,333 |  |
| SE aligned 1 time (%) | 37.00 | 8.72 | 12.38 |
| SE aligned >1 time (%) | 14.68 | 0.32 | 0.02 |
| Total # of PE reads |  | 1,447,813,431 |  |
| PE p. aligned 1 time (%) | 36.78 | 7.41 | 9.48 |
| PE p. aligned >1 time (%) | 8.90 | 0.20 | 0.01 |
| PE improperly aligned (%) | 2.67 | 0.54 | 0.82 |

SE: Single-end; PE: Paired-end; p.: Properly; Bowtie2 were run with "-L 27".

## ACKNOWLEDGEMENTS

We thank S.M. Yiu, C.M. Leung and Y. Peng for the detailed explanation about IDBA-UD. We also thank C. Titus Brown for providing the open evaluation with the *E. coli* data (Table 1).
*Conflict of Interest*: None declared.